%% file: paper.tex
\documentclass{pasj00}
\draft

\begin{document}
\SetRunningHead{Author(s) in page-head}{Running Head}
\Received{2006/12/31}%{yyyy/mm/dd}
\Accepted{2007/2/10}%{yyyy/mm/dd}

\title{Long Term Simulations Of Astrophysical Jets;\\ 
Energy Structure and Quasi-Periodic Ejection}

%%% begin:list of authors
\author{Ahmed \textsc{Ibrahim}}
\affil{Kwasan Observatory, Kyoto University, Yamashina-ku, Kyoto 607-8471}
\email{aaeakf1@kwasan.kyoto-u.ac.jp}

\and
\author{Kazunari {\sc Shibata}}
\affil{Kwasan Observatory, Kyoto University, Yamashina-ku, Kyoto 607-8471}
\email{shibata@kwasan.kyoto-u.ac.jp}

%%% end:list of authors

%%% Please use the following style in case that sorting by 
%%% affilation is impossible. 
%
% \author{%
%   D-Firstname \textsc{D-Familyname}\altaffilmark{1}
%   E-Firstname \textsc{E-Familyname}\altaffilmark{1,2}
%   and
%   F-Firstname \textsc{F-Familyname}\altaffilmark{2}}
% \altaffiltext{1}{Address of Institute}
% \email{ddddd@xxx.xxx.xx.xx}
% \email{eeeee@xxx.xxx.xx.xx}
% \altaffiltext{2}{Address of Institute}

%% `\KeyWords{}' always has to be placed before `\maketitle'.
 %Do NOT move this preamble from here!

\maketitle

\begin{abstract}
We have performed self-consistent 2.5-dimensional nonsteady MHD numerical
 simulations of jet formation as long as possible, including the dynamics of
 accretion disks. Although the previous nonsteady MHD simulations for
 astrophysical jets revealed that the characteristics of nonsteady jets
 are similar to those of steady jets, the calculation time of these
 simulations is very short compared with the time scale of observed jets. 
Thus we have investigated long term evolutions of mass accretion rate, mass outflow
 rate, jet velocity, and various energy flux. 
We found that the ejection of jet is quasi-periodic. The period of the ejection
 is related to the time needed for the initial magnetic filed to be twisted to
 generate toroidal filed
\[
T_{ejection} \propto \frac{1}{V_A} \propto \frac {1}{B} \propto E^{-\frac{1}{2}} _{mg}.
\]
 We compare our results with both
 the steady state theory and 
previous 2.5-dimensional nonsteady MHD simulations. Then it is found that time
 averaged velocity of jets ($V_{jet \hspace{0.1cm}\mathrm{ave}}$)
 are $\sim 0.1V_{\mathrm{K}}$ 
and $\sim 0.1V_{jet \hspace{0.1cm}\mathrm{max}}$, where $V_{\mathrm{K}}$ is the
 Keplerian velocity at $(r,z)=(1,0)$ and $V_{jet \hspace{0.1cm}\mathrm{max}}$
 is
 the maximum velocity of jet. Nevertheless, the characteristics of 
our simulations are consistent with those of 
steady solution and previous short term simulations 
in that the dependences of the time averaged velocity
 $V_{z \hspace{0.1cm}\mathrm{ave}}$ 
and mass outflow rate $\dot{M}_{w \hspace{0.1cm}\mathrm{ave}}$ 
on the initial magnetic field strength are approximately 
\[
\dot{M}_{w \hspace{0.1cm}\mathrm{ave}} \propto B^{0.32}\hspace{0.5cm}
\mathrm{and} \hspace{0.5cm} V_{jet \hspace{0.1cm}\mathrm{ave}} \propto {B_0}^{0.3}.
\]

\end{abstract}

\newpage

%##########################################################
\section{Introduction}
%#########################################################

Astrophysical jets have been observed in young stellar 
objects (YSOs) (e.g., \cite{Fuk1993}; \cite{Ogu1995}; 
\cite{Bur1996}; \cite{Bac1996}; \cite{Rei2002}; \cite{Cur2006}), active galactic nuclei 
(AGNs) (e.g., \cite{BrP1984}; \cite{Bir1995}; \cite{Jun1999}; \cite{JiH2003};
 \cite{Mat2005}), and some X-ray binaries (XRBs) (e.g., \cite{Mar1984}; \cite{Mir1994};
\cite{Tin1995}; \cite{Kot1996}; \cite{Mig2006}). Although the acceleration and
 collimation mechanisms of these jets are still not well understood, these objects
 are believed to have accretion disks in their central regions.

The standard model of a jet-disk system has been an original idea of Blandford 
\& Payne (1982).
 Energy and angular momentum are removed magnetically from the 
accretion disk by field lines anchored to the disk surface and extending to 
large distances. Blandford \& Payne (1982) showed that a centrifugally driven
 outflow of matter from the disk is possible, if the angle between field line
 and disk is less than $60^\circ$. They discussed self-similar solutions of
 the steady and axisymmetric MHD equations and the possibility of such
 acceleration and collimation of the flow from a cold Keplerian disk.
 After their work, many authors studied MHD models of jet formation 
from accretion disks (e.g., \cite{PuN1986}; \cite{Sak1987}; \cite{Lov1991};
 \cite{CoL1994}; \cite{NaS1994}; \cite{FeC1996}) based on the theory of steady and
 axisymmetric MHD winds, which was first developed by  Weber \& Davis (1967)
 for the solar wind.
 Cao \& Spruit (1994) examined the mass outflow rate of magnetically driven jets by studying 
the solution that passes through the slow magnetosonic 
point (see also \cite{Li1995}). They confirmed that the inclination 
angle of the field line is very important for achieving a high mass outflow rate.

Kudoh \& Shibata (1995, 1997a) studied one-dimensional 
steady magnetically driven jets along a fixed poloidal field 
line for a wide range of parameters, assuming the shape of 
that field line. They found that the steady solutions generally 
can be classified into two branches: (1) magneto-centrifugally 
driven jets, when the magnetic field is strong, in that case they
 found that the jet velocity $(V_{jet})$ and mass outflow rate
 $( \dot{M}_{w \hspace{0.01cm}}= \frac{dM_{w}}{dt})$ depend on the
 magnetic energy $(E_{mg})$ as $V_{jet \hspace{0.01cm}}
 \propto {B_0}^{1/3}$, $\dot{M}_{w \hspace{0.01cm}} \sim constant$
 and (2) magnetic pressure-driven, when the magnetic field is weak, 
in that case $V_{jet \hspace{0.01cm}}
 \propto {B_0}^{1/6}$, $\dot{M}_{w \hspace{0.01cm}} \propto {B_0}^{0.5}$.
 In the centrifugal case, the main acceleration occurs before the 
flow speed exceeds the Alfv\'en speed, while in the latter case 
most of the acceleration occurs after the flow speed becomes greater than
 the Alfv\'en speed. This mass flux scaling law was confirmed by Ustyugova et al. (1999).

 In most theoretical models of jets from accretion
 disks, however, accretion disks are treated as boundary conditions. Accretion
 disks only play a role of supplying energy and mass to the jets, and neither
 accretion flow nor internal structure of disks are considered in these models.
 Since the disk itself is not treated,
 such simulations may last over hundreds of Keplerian periods. This idea
 was first applied by Ustyugova et al. (1995). Extending this work, Romanova
 et al. (1997) found
 a stationary final state of a slowly collimating disk
 wind in the case of a spilt-monopole initial field structure after 100
 Keplerian periods. Ouyed \& Pudritz (1997a, 1997b) presented time-dependent
 simulations of the jet formation from a Keplerian disk. For a certain
 (already collimating) initial magnetic field distribution, a stationary state
 of the jet flow was obtained after about 400 Keplerian periods of the inner
 disk with an increased degree of collimation. Ouyed et al. (2003) investigate the
 problem  of jet stability and magnetic collimation extending the axisymmetric
 simulations to 3D.

On the other hand, Uchida \& Shibata (1985) and Shibata \& Uchida (1987, 1989, 1990) 
performed time-dependent, two-dimensional axisymmetric 
(2.5-dimensional) MHD numerical simulations of magnetically driven jets from 
accretion disks. They solved the interaction between a geometrically 
thin rotating disk, including the dynamics of the disk,
 and a large-scale magnetic field that was initially
 uniform and vertical . Shibata \& Uchida (1986) investigated the detailed properties of these
 jets. They found that (1) the velocity of the jet was typically of order of the disk's
 Keplerian velocity and (2) it increased with increasing magnetic field 
strength in a manner similar to the scaling law of  Michel's (1969) solution. 
 Matsumoto et al. (1996) carried out 2D MHD simulations of a torus threaded by poloidal
 magnetic fields and found that (1) the jet velocities 
were again comparable to the Keplerian velocity and (2) the 
mass outflow rate of the jet increased with the strength of the initial magnetic field.
 Kudoh et al (1998) studied the formation mechanism of jets
 from geometrically thick disks and the dependence of the initial magnetic
 field strength $(B_0)$ in detail by performing self-consistent 2.5-dimensional
, nonsteady, ideal MHD simulations including the dynamics of the disk. They 
found that the ejection point is determined by the effective potential
 resulting from the gravitational and centrifugal forces along the field 
lines (\cite{BlP1982}) and also that the velocity and mass outflow rate are
 consistent with those predicted by the steady theory (\cite{KuS1995} and \cite{KuS1997a}).
 Kato et al (2002) performed 2.5-dimensional, axisymmetric, ideal
 MHD simulations of jets from geometrically thin disks for Keplerian 
and sub-Keplerian cases over a wide range of initial magnetic field 
strengths. Kigure \& Shibata (2005) investigate the problem of jet formation
 and stability by using 3-dimensional MHD simulations. To investigate the
 stability of the MHD jet, they introduce a perturbation to the accretion
 disk with a nonaxisymmetric sinusoidal or random fluctuation of rotational 
velocity. In both perturbation cases, a nonaxisymmetric structure with $m=2$
 appears in the jet, where m is the azimuthal wavenumber. They conclude that
 this structure seems to originate in the accretion disk.

On the other hand, the acceleration and collimation of the jet have been
 studied in the steady (e.g. \cite{Sau2004}; \cite{BoT2005}).
In recent years, many simulations taking other physical processes into
 consideration, e.g. the magnetic diffusion ( \cite{Kuw2000}, \cite{Kuw2005}; 
\cite{FeV2002}; \cite{CaK2002}, \cite{CaK2004}), the dynamo process in
 the accretion disk (\cite{von2003}), and the radiation force 
( \cite{Pro2003}), have been performed. Koide et al. (1999) showed that
 a magnetically driven jet in the general relativistic MHD simulation has
 characteristics similar to those of the nonrelativistic MHD jet
 (\cite{ShU1986}). In addition to these studies which consider the
 initial magnetic field as large scale, several papers were considering the
 evolution of a stellar magnetic dipole in interaction with a diffusive
 accretion disk. Hayashi et al. (1996) observed magnetic reconnection and 
evolution of X-ray flares during the first rotational periods in 
their numerical simulations.

We note that all calculations including the treatment of the disk structure
and the ideal MHD, (e.g. \cite{ShU1986}, \cite{ShU1987}, \cite{Shi1989}, \cite{ShU1990};
  \cite{Mat1996}; \cite{Kud1998}; \cite{Kat2002}; \cite{Kud2002} and \cite{KiS2005}), 
could be performed only for $1-2$ Keplerian periods of inner disk. At this
 point, we emphasize that the observed kinematic time scale of protostellar
 jets can be as large as $10^3-10^4$ yrs, corresponding to
 $5\times 10^4-5\times 10^5$ stellar rotational periods
 (and inner disk rotations)! For example, proper motion measurements
 for HH30 jet (\cite{Bur1996}) give a knot velocity of
 about $100-300 km s^{-1}$ and a knot production rate of about 0.4 knot
 per year. Assuming a similar jet velocity a long the whole jet extending
 a long 0.25 pc (\cite{Lop1995}), the kinematic age is about 1000 yrs

In order to have access to the observed time scale of jets and to know whether
 the jet formation becomes quasi-steady state, we performed long term
 2.5-dimensional MHD simulations of jets.

We solve the dynamics of the disk as in Shibata \& Uchida (1986), Kudoh et al. (1998) and
 their following studies.
 We also want to know whether the time averaged physical
 quantities have the same characteristics as those in the steady model and
 previous simulations (\cite{Kud1998}) and (\cite{Kat2002}). Our
 calculation time is about 20 times longer than those of previous
 simulations. Also, the simulation box is large enough to minimize  the
 effect of top and side boundary conditions
%#####################################################
\section{Numerical Method}
\subsection{Assumptions}
%#####################################################
Our simulations make the following assumptions: (1)
axial symmetry around the rotational axis $(\partial /\partial \varphi = 0)$, 
including the azimuthal components of a velocity $(v_{\varphi})$ and a 
magnetic field $(B_\varphi)$ (i.e., 2.5-dimensional approximation); (2) 
ideal MHD (i.e., Ohmic diffusivity and/or ambipolar diffusion 
are assumed to be negligible); (3) an inviscid perfect gas 
with a specific heat ratio of $\gamma = 5/3$; (4) a point-mass gravitational 
potential only, with disk self-gravity neglected.
\subsection{Basic Equations }
The basic equations we use are the 2.5-dimensional 
ideal MHD equations in cylindrical coordinates $(r, \varphi, z)$ :

\begin{equation}
\frac{\partial \rho}{\partial t} + v_r \frac{\partial \rho}{\partial r} + 
v_z \frac{\partial \rho}{\partial z} = -\rho \Bigl( 
\frac{1}{r}\frac{\partial}{\partial r}(r v_r) + \frac{\partial v_z}{\partial z} \Bigr), 
\end{equation}
\begin{eqnarray}
\frac{\partial v_r}{\partial t} + v_r\frac{\partial v_r}{\partial r} + 
v_z \frac{\partial v_r}{\partial z} = - \frac{1}{\rho}\frac{\partial p}{\partial r} + \frac{{v_\varphi}^2}{r} - 
\frac{\partial \Psi}{\partial r} - \frac{1}{4 \pi \rho} \nonumber \\
&& \hspace{-4.0cm} \times \Bigl[ B_z \frac{\partial B_z}{\partial r} 
+ \frac{B_{\varphi}}{r} \frac{\partial}{\partial r}(rB_\varphi) \Bigr] + 
\frac{B_z}{4\pi \rho}\frac{\partial B_r}{\partial z},
\end{eqnarray}
\begin{equation}
\frac{\partial (rv_\varphi)}{\partial t} + v_r \frac{\partial (rv_\varphi)}{\partial r} + 
v_z \frac{\partial (rv_\varphi)}{\partial z} = \frac{1}{4 \pi \rho} 
\Bigl[ B_r \frac{\partial}{\partial r}(rB_\varphi) + B_z \frac{\partial}{\partial z}(rB_\varphi) \Bigr],
\end{equation}
\begin{eqnarray}
\frac{\partial v_z}{\partial t} + v_r\frac{\partial v_z}{\partial r} + 
v_z \frac{\partial v_z}{\partial z} = 
- \frac{1}{\rho}\frac{\partial p}{\partial r} + 
\frac{\partial \Psi}{\partial r} - \frac{1}{4 \pi \rho} \nonumber \\
&& \hspace{-3cm} \times \Bigl( B_r \frac{\partial B_r}{\partial z} 
+ B_{\varphi} \frac{\partial B_\varphi}{\partial z} \Bigr) + 
\frac{B_r}{4\pi \rho}\frac{\partial B_z}{\partial r},
\end{eqnarray}
\begin{equation}
\frac{\partial e}{\partial t} + v_r\frac{\partial e}{\partial r} + 
v_z\frac{\partial e}{\partial z} = -\frac{p}{\rho}\Bigl[ 
\frac{1}{r}\frac{\partial}{\partial r}(rv_r) + \frac{\partial v_z}{\partial z} \Bigr],
\end{equation}
\begin{equation}
e = \frac{p}{(\gamma -1) \rho},
\end{equation}
\begin{equation}
\frac{\partial B_r}{\partial t} = -\frac{\partial}{\partial z}(v_z B_r - v_r B_z ),
\end{equation}
\begin{equation}
\frac{\partial B_{\varphi}}{\partial t} = \frac{\partial}{\partial z}
(v_{\varphi}B_z - v_z B_{\varphi}) - \frac{\partial}{\partial r}
(v_r B_{\varphi} - v_{\varphi}B_r),
\end{equation}
\begin{equation}
\frac{\partial B_z}{\partial t} = \frac{1}{r}\frac{\partial}{\partial r}
\Bigl[ r(v_z B_r - v_r B_z) \Bigr],
\end{equation}
where $\Psi = -GM/(r^2 + z^2)^{1/2}$ is the gravitational potential, 
$G$ is the gravitational constant, and $M$ is the mass of a central object.
Other variables are summarized in Table 1.

%###############################################################
\subsection{Initial Conditions }
%###############################################################

As an initial condition, we assume an equilibrium disk rotating in a central 
point-mass gravitational potential (\cite{Mat1996}). 
Exact solutions for these conditions can be obtained under the simplifying 
assumptions for the distribution of angular momentum and pressure 
(e.g., \cite{Abr1978}):
\begin{equation}
L = L_0 r^a,
\end{equation}
\begin{equation}
p = K \rho^{1+1/n}.
\end{equation}

Then the distribution of material in the disk is given by 

\begin{equation}
\Psi = - \frac{GM}{(r^2 + z^2)^{1/2}} + \frac{1}{2(1-a)}{L_0}^2r^{2a-2}
+( n + 1 )\frac{p}{\rho} =\mathrm{constant}.\label{12}
\end{equation}

We use $a = 0.45$ and $n = 3$ throughout this paper. 
These are the different parameters from the previous similar simulations
(e.g.,\cite{Kud1998}, \cite{Kat2002}). Our parameters  make the disk thicker
 and wider than that of Kato et al. (2002), 
mainly for both numerical convenience and long term simulation. 
The mass distribution outside the disk is assumed to be 
that of a uniformly high temperature corona in hydrostatic 
equilibrium without rotation. The density distribution in 
hydrostatic equilibrium is 
\begin{equation}
\rho = \rho_c \mathrm{exp} \biggl\{ \alpha \biggl[ \frac{r_0}{(r^2 + z^2)^{1/2}} -1 \biggr] \biggr\}.
\end{equation}
where $r_0$ is a radius defined later (see Table 1), 
$\alpha = (\gamma {V_{k0}}^2/{V_{sc}}^2), V_{sc}$ is the sound velocity in the corona, 
$V_{k0} = (GM/r_0)^{1/2}$ is the Keplerian velocity at radius $r_0$. 
We use $\alpha = 1$ and $\rho_c/\rho_0 = 10^{-3}$ throughout this paper, 
where $\rho_0$ is a density defined later (see Table1).

For simplicity, the initial magnetic field is assumed to be 
uniform and parallel to the axis of rotation 

\begin{equation}
B_z = B_0 = \mathrm{constant}, \hspace{0.5cm} B_r = B_{\varphi} = 0.
\end{equation}

Indeed, the presence of a large-scale magnetic field in accretion-disk-jet 
systems is observed in AGN jets and protosteller jets (\cite{Ray1997}; \cite{Pus2005};
 \cite{Gab2004}; \cite{Wou2006}) and is assumed theoretically in jet 
launching models (e.g. \cite{CaK2004}, and references therein)
%#############################################
\subsection{Boundary Conditions}
%#############################################
 On the axis $(r = 0)$ we assume a boundary condition that is symmetric for
 $\rho, p, v_z,$ and $B_z$  while $v_r, v_{\varphi}, B_r,$ and $B_{\varphi}$
 are antisymmetric.

The side, top and bottom surfaces are free boundaries. 
In order to avoid a singularity at the origin, the region around 
$r = z = 0$ is treated by softening the gravitational potential as 
\begin{equation}
\Psi = 
\Biggl\{
\begin{array}{ccc}
-GM/(r^2+z^2)^{1/2} 
&\mathrm{for}& \epsilon < (r^2 + z^2)^{1/2}, \nonumber \\
-GM{1/\epsilon - [(r^2 + z^2)^{1/2} - \epsilon]/\epsilon^2} 
&\mathrm{for}& 0.5\epsilon < (r^2 + z^2)^{1/2} \leq 
\epsilon, \nonumber \\
-1.5GM/\epsilon 
&\mathrm{for}&  (r^2 + z^2)^2 \leq 
0.5 \epsilon.
\end{array}
\end{equation}

We use $\epsilon = 0.2r_0$ throughout this paper.

These boundary conditions are different from those  MHD 
simulations of jets from accretion disks performed by Ouyed et al (1997a),
 Romanove et al. (1997), Meier et al. (1997), Ustyugova et al. (1999),
 and Pudritz et al.(2006).
 The main difference is the condition on the equatorial plane. They assumed an inflowing
 fixed boundary condition (i.e., the angular momentum is continually injected 
from the boundary). Therefore, their numerical simulations did not allow disk accretion
 due to magnetic braking nor growth of magneto-rotational instability.
%############################################
\subsection{Numerical Schemes}
%###########################################
The numerical schemes we use are the cubic-interpolated pseudo-particle (CIP)
 method (\cite{YaA1991}; \cite{Yab1991}) and the method of
 characteristics-constrained transport (MOCCT) (\cite{EvH1988}; \cite{StN1992}). 
The magnetic induction equation is solved using 
MOCCT and the others using CIP. This is the two-dimensional 
version of the scheme used by  Kudoh \& Shibata (1997b), and Koide et al. (1999)

%###########################################
\subsection{Parameter} 
%##########################################

We normalize the physical quantities with their initial value 
at $(r,z) = (r_0,0)$ taking $r_0 = ({L_0}^2/GM)^{1/(1-2a)}$ so that 
the initial density of the disk has maximum value of $\rho_0$ at 
$(r,z) = (r_0,0)$. The normalized unit for each variable is 
summarized in Table 1.
There are two nondimensional parameters:

\begin{equation}
E_{\mathrm{th}} = \frac{{V_{s0}}^2}{\gamma {V_{K0}}^2} \mathrm{,},
\end{equation}
\begin{equation}
E_{\mathrm{mg}} = \frac{{V_{A0}}^2}{{V_{K0}}^2} \mathrm{,},
\end{equation}
where $V_{s0} = (\gamma p_0/\rho_0)^{1/2}$, $V_{A0} = B_0/(4\pi \rho_0)^{1/2}$, and 
$p_0$ is the initial pressure at $(r,z) = (r_0, 0)$. 
 The initial parameters we use in this paper are summarized in Table 2. 
We use $E_{\mathrm{th}} = 0.018$ throughout this paper.
When $E_{\mathrm{th}} = 0.018$ and $\alpha = 1$, the temperature of corona 
is about $10^{2}$ times 
greater than that of the disk at $(r,z) = (r_0,0)$.
In this paper, we adopt $L=L_0r^{0.45}$ $(L_0=1.00)$ that makes 
the disk thicker and 
wider than that of  Kato et al. (2002), 
mainly for both numerical convenience and long term simulation: 
the internal structure of a thick disk is better resolved than that of a thin disk,
and much matter in the disk can make the time of simulations longer than that of
  Kudoh et al. (1998). Of course, many accretion disks in YSOs, 
AGNs, and XRBs are expected to be geometrically thin and 
nearly Keplerian.  Kato et al. (2002) also performed simulations for geometrically thin
 disk case. However they calculated only one or two orbits just the same as
  Kudoh et al. (1998). The minimum grid size is $0.05r_0$ in r-direction and $0.01r_0$
 in z-direction. The maximum grid size in r-direction is  $0.1r_0$ and  $0.4r_0$ in
 z-direction, and the number of grid points used in this paper is $269 \times 639$.
 The grid spacing is uniform for $r/r_0 < 1$ and $z/r_0 < 1$ and stretched in
 $r$ and $z$ for $r/r_0 > 1$ or $z/z_0 > 1$. The size of the computational domain is
 $(R_{\mathrm{min}}/r_0 \sim R_{\mathrm{max}}/r_0\mathrm{,} \hspace{0.1cm}
 Z_{\mathrm{min}}/r_0 \sim Z_{\mathrm{max}}/r_0) = (0.0 \sim 24.8 \mathrm{,} \hspace{0.1cm} -65.75 \sim 65.75)$. 

%####################################################
\section{Numerical Results}
%####################################################
%#############################
\subsection{Typical Case}
%#############################
Figures \ref{1}, \ref{2} and \ref{3} shows the time evolution of the density 
distribution 
(color scales, top panels), temperature distribution 
(color scales, bottom panels) 
 $(T \equiv \gamma p/\rho)$, poloidal magnetic field lines, and poloidal 
velocity (hereafter, the variables are expressed in nondimensional form)
for model 1,4 and 6 respectively. Time $t = 2\pi \simeq 6.28$ corresponds
 to one Keplerian orbit at $(r,z) = (1,0)$. Evolutions for all models look 
like similar each other. 

Now we show the case of model 4 (see Figure \ref{2}).
In the early stage of evolution, a torsional Alfv\'en wave is generated at 
the disk surface and propagates up into the corona $(t=13)$. Since this wave
 extracts angular momentum, the rotating disk begins to fall into the central 
region. The surface layer of the disk falls faster than the 
equatorial part. This is the avalanche-like accretion that was studied
 by Matsumoto et al. (1996). Because only a small fraction of the accreted
 matter is ejected into the bipolar direction due to Lorentz force in the
 relaxing twist, both density and pressure of the inner region increase.
 Within this region, angular momentum is transferred from the high to the low-density
 parts. The cold material on the disk surface is ejected as a jet $(t=35)$. At $t=82$
 due to the magnetorotational instability the channel flow becomes clear. 
Initially both gas pressure and magneto-centrifugal force drives
 and accelerates the outflow (below the Alfv\'en surface). After that,
 when the toroidal field generated as a result of the differential rotation 
of the accretion disk is accumulated, the acceleration is  due to the magnetic
 pressure gradient, $\partial (B^2_\varphi/8\pi)/\partial z$. To illustrate
 the twist level of the magnetic field lines, we show in figure \ref{4}
 the time variation of the ratio
of the toroidal magnetic field, $B_\varphi$, to the poloidal magnetic field
, $B_p$, along $r=0.225$. Figure \ref{4} illustrates that in case of
 initial weak magnetic field the twisting field $B_\varphi$ become more 
dominant and both of
 the Alfv\'en point and the slow point become near to the accretion disk
 (\cite{PeP1992} and \cite{KuS1997a}). In both the weak and strong initial 
magnetic field a strong twist appears at t=15 and propagates outward when the
 jet is ejected.

The outflow  consists of both the material that is initially in the corona and
the material from the disk. Channel flow continues to grow in the 
disk and jets are ejected continuously and intermittently $(t=82)$. 
Jet ejection
and accretion still continue at the last stage of evolution in
our simulations $(t=115)$. We can also see that magnetic field lines 
entwine each other and that the magnetic turbulent flow develops. We can
see that the magnetic islands are created by magnetic reconnections in the disk
. Since we assumed ideal MHD, the magnetic reconnection is caused by numerical
 diffusion. Numerical diffusion is inevitable for finite difference numerical
 schemes, even though we do not include the magnetic diffusivity explicitly.
 The time evolution is similar to that in Matsumoto et al. (1996) and Kudoh et al. (1998).

The avalanche-like flow is caused by a physical mechanism 
similar to that of the magnetorotational instability 
(\cite{BaH1991}; \cite{HaB1991}). Note that although the disk is initially rotating
 in our model, the corona is not. The discontinuity of azimuthal velocity at the 
interface between them generates torsional  Alfv\'en waves 
propagating into both the corona and the disk. Since the 
 Alfv\'en wave behaves like a large amplitude perturbation, it 
triggers the magnetorotational instability in the disk. 
The wavelength of this instability is determined by the most unstable
 wavelength $(\lambda)$ for the magnetorotational instability, $\lambda \sim 2\pi
 V_A/\Omega$, so the base height of the avalanche depends on the initial 
magnetic field strength.

%################################################################
\section{Time Evolutions of Some Physical Quantities of Jets}
%################################################################
One of the most important purpose of this study is to clarify whether 
the jet ejection becomes steady state. In figures \ref{5},\ref{6} and \ref{9}
we show the time evolutions of the mass outflow rate $(\dot{M_w})$, 
the mass accretion rate $(\dot{M_a})$, and the toroidal magnetic field energy 
$(E_{mgt})$ for  model 1 $(E_{mg}=2\times 10^{-4})$, model 4 
$(E_{mg}=2\times 10^{-5})$ and model 6 $(E_{mg}=2\times 10^{-6})$.
 $\dot{M_w}$ is defined by 
\begin{equation}
\dot{M_w} = \pi \Biggl( \int_{0}^{1} \rho v_z {\scriptstyle (r,z=z_p)} r dr 
- \int_{0}^{1} \rho v_z {\scriptstyle (r,z=-z_p)} r dr \Biggr).\label{18}
\hspace{1cm} 
\mathrm{at\hspace{0.2cm}z} = 4.
\end{equation}
where $z_p$ is z-coordinates for jets to pass through. 
The mass accretion rate for various models. 
$\dot{M_a}$ is defined by 
\begin{equation}
\dot{M_a} = - \pi \int_{-1}^{1} \rho v_r{\scriptstyle (r=1.0,z)} r dz.  
\end{equation}
It is not easy to define $(\dot{M_w})$ and $(\dot{M_a})$, 
because of following reasons: 
(1) ejection and accretion are very nonsteady, and sometimes the ejected matter
falls back onto the disk, 
(2) when initial magnetic field is weak, the jet ejection point goes away 
from the equatorial plane after long time simulation. Kudoh et al. (2002) 
and Kuwabara et al. (2000) found that mass accretion and mass ejection take
 place intermittently in the case without perturbation in the disk. The 
simulation time of both of them ranged from one orbit in case of Kuwabara et al. (2000)
and three orbits in case of Kudoh et al. (2002). Our simulations in this 
paper more than 20 orbits for different models. From Figure \ref{5}, it is clear that 
the mass ejection flux is still intermittent until the last stage of evolution.
 The intermittent ejection is clear in all models in the figure \ref{5} but there are
 some differences between them. The first one, in case of strong magnetic field the 
intermittency is almost similar at the beginning of the simulation to that at
 the end of simulation. But in case of weak field, as the simulation goes on, 
the ejection mass flux becomes very nonsteady. Not only 
the intermittency increases in case of weak magnetic field but also the 
absolute amount of mass flux increases. But in case of strong magnetic field
 we notice the opposite. Figure \ref{6} shows the mass accretion rate
 for models 1, 4 and 6. The mass accretion, like mass ejection, is intermittent
 until the last 
stage of our simulations. In both weak and strong initial magnetic field the
 mass accretion is intermittent. The general trend of intermittency of the
 mass accretion is similar to the intermittency of mass ejection. That 
similarity is more clearer for example in model 4 than other models.
 In model 4 the absolute value 
of mass accretion at the beginning of simulation is smaller than that at the end. We
notice the trend in the mass ejection, also the same trend happen in the other models.

Because of the mass accretion rate is highly variable, plotting it
at one radius in time shows very little of the overall characters of the
 accretion. We therefore show values that are averaged over space or time, in 
order to obtain a better understanding of the overall accretion within the
 disk.

 Figure \ref{7} shows the accretion rate against time,
 averaged between $r= 0.52$ 
and $r=10$. This region is chosen since the accretion has little value at 
larger radii. This shows that even though the accretion rate is highly
 intermittent, it is predominantly positive. The behavior of the accretion 
is more complex in case of weak initial magnetic field after 10 orbits, 
because the evolution of the disk take a long time until it becomes more
 turbulent. The accretion, in case of strong initial magnetic field, is
 positive during the first 5 orbits. During that time, part of the subtracted
 angular momentum is ejected as a jet. The other part is transported at large 
distances by the magnetic stresses. This is exactly what we see in
 Figure \ref{8} , 
which shows the accretion rate as a relation of radius of the disk, averaged 
over the first 10 orbits. Accordingly, we see that the accretion is positive 
in the inner part of the disk and the accretion is negative beyond the
 radius 2-4. The negative accretion continue about 5 orbits.
 The accretion becomes positive again because the continuous ejection of 
the jet which means continuous subtraction of the angular momentum. 

Figure \ref{9} shows the toroidal magnetic energy for model 1, 4 and 6.
The toroidal magnetic field energy $E_{mgt}$ for various models is defined by 
\begin{equation}
E_{mgt} = \pi \Biggl( \int_{0}^{1} \frac{{B^2}_{\varphi}{\scriptstyle (r,z=z_p)}}{8\pi} rdr  
+ \int_{0}^{1} \frac{{B^2}_{\varphi}{\scriptstyle (r,z=-z_p)}}{8\pi} rdr  \Biggr)\bigg/\pi\int_{0}^{1} 2rdr. \label{19}
\end{equation}
The intermittency is clear and the trend is also similar to mass ejection. 
Hence we try to find the relation between mass accretion, ejection and
 magnetic field. The ejections are intermittent and seem to
 have periods for all models, and the periodicity seems  
different in the models with different initial magnetic field.
Next we check the times of ejection peaks for different models and study the 
relation between that peak times with the time needed for the toroidal field 
to be accumulated after that untwisted in vertical direction carrying the
 mass flux. From the initial conditions equation \ref{14},  $B_z = B_0 = \mathrm{constant}, \hspace{0.5cm} B_r = B_{\varphi} = 0$. With the rotation of the
 disk by angular velocity $\Omega$, the toroidal field $B_\varphi$ is generated
 from $B_z$.
\begin{equation}
\frac{\partial B_\varphi}{\partial T} \approx \Omega B_z.
\end{equation}
By integration w.r.t. time T then 
\begin{equation}
B_\varphi \approx \Omega B_z T.\label{22}
\end{equation}
The ejection occurs when the magnetic energy equals the rotational energy, then
\begin{equation}
\frac{B^2_\varphi}{8\pi} \approx \frac{1}{2} \rho V^2_\varphi \approx \frac {1} {2} \rho V^2_k.\label{23}
\end{equation}
Combining Eqs. \ref{22} and \ref{23}, yields
\begin{equation}
\frac{ \Omega^2 B^2_z T^2}{8\pi} \approx \frac{1}{2} \rho V^2_\varphi \approx \frac {1} {2} \rho V^2_k,
\end{equation}

\begin{equation}
T_{ejection} \approx \frac{\sqrt{4\pi \rho}}{B_z} \frac{V_k}{\Omega} \approx \frac{1}{V_A} \frac{V_k}{\Omega},\label{25} 
\end{equation}

\begin{equation}
T_{ejection} \propto \frac{1}{V_A} \propto \frac {1}{B} \propto E^{-\frac{1}{2}} _{mg},\label{26}
\end{equation}
where $V_A$ is the  Alfv\'en speed, $B$ is the initial magnetic field strength and
 $T_{ejection}$ is the time corresponding to the peak mass ejection rate as 
shown in Figure \ref{10}. Figure \ref{11} shows the relation between
 the initial magnetic
 field strength and the average time interval between the peaks of the 
mass ejection rate of models
 1, 3, 4 and 6. Here the squares correspond to our numerical values and 
the solid line is the best fit to them with the following dependence 
\[
T_{average, eje}\propto E^{-0.3} _{mg}.
\] 
The dotted line corresponds to the analytical relation equation \ref{26}.
We notice that the dependence of the ejection time on the initial magnetic
 field $(E_{mg})$ is in a good agreement with our analytical expectation
 equation \ref{26}. The small deviation from the analytical relation comes from our
 assumption that  both the
 magnetic field and density will remain constant during the simulation. But
 the real situation is that both the density and magnetic field increase 
with time during the simulation. By taking the time evolution
 of both density and magnetic into account (inside the disk),
 the equation \ref{25} becomes
\begin{equation}
T_{ejection}\propto \frac{\sqrt{\rho}}{B_z}\propto \frac{1}{V_A}. 
\end{equation} 
Both the density and $B_z$ component of the 
magnetic field increase steeply in case of strong initial magnetic field.
 On the other hand, the density almost remains  constant in case of initial
 weak magnetic field while $B_z$ increases. Then as $B_z$ increases with time 
the ejection time becomes shorter than if we consider constant initial
 magnetic field. But the time evolution of the density will
 make an opposite effect in that case of strong initial magnetic field i. e.
 the effect of the increase of $B_z$ will be weakened by the effect of the 
 increase in $\rho$. On the other hand, in the case of initial weak magnetic
 field, the effect of the increase of the density is too small to weaken the
 effect of the increase of magnetic field. Hence the evolution effect of 
 $B_z$ is more prominent in case of weak initial magnetic field.
 So we think that the evolution of both density and magnetic field $B_z$ may
 explain the discrepancy between the analytical dependence of the ejection
 time and the initial magnetic field. Consequently, when we consider the average
 magnetic field instead of the initial magnetic field in Figure \ref{11}a,
 the agreement is more better as we can see from Figure \ref{11}b.
 In that case the 
best fit is
\[
T_{average, eje}\propto E^{-0.36} _{mg}.
\]
%##########################################################################
\section{Time averaged Velocities and Mass Outflow Rates of Jets As a Function of Magnetic Energy}
%##########################################################################
There is also an interesting question regarding magnetically driven 
jets: how the velocities and mass outflow rates depend on the strength of the
 magnetic field. Figure \ref{12}a shows the time averaged jet
 velocities
 ($V_{jet \hspace{0.1cm} \mathrm{ave}}$) as a function of the initial
 magnetic field $E_\mathrm{mg}$. 
$V_{jet \hspace{0.1cm} \mathrm{ave}}$ is defined by 

\begin{equation}
V_{jet \hspace{0.1cm} \mathrm{ave}} = 
\int_{0}^{T_e} \int_{0}^{4} v_z {\scriptstyle (r,z=z_p)} dr dt 
\bigg/ 10 T_{orbit}, \hspace{1cm}. 
\end{equation}
where $T_e$ is the end of the simulation time, $T_{orbit}$ is the time
 of one disk rotation and $10T_{orbit}$ is the sum of the 
time over which  the speed of the jet is averaged. 
The figure shows that \textit{the jet velocity is} $10^{-1}$ 
\textit{smaller than that of order the Keplerian speed} 
for a wide range of $E_{\mathrm{mg}}$ in the disk and 
that its dependence on $E_{\mathrm{mg}}$ as shown in  Figure
 \ref{12}a is approximately.
\[
V_{jet \hspace{0.1cm} \mathrm{ave}} \propto {E_{\mathrm{mg}}}^{0.15}.
\]

Figure \ref{12}b shows the maximum jet velocities
 ($V_{jet \hspace{0.1cm} \mathrm{max}}$) as a function of the initial
 magnetic field $E_\mathrm{mg}$.
The sold line shows
\[
V_{jet \hspace{0.1cm} \mathrm{max}} \propto {E_{\mathrm{mg}}}^{0.17}.
\]

Figure \ref{12}c and \ref{12}d show the time averaged jet velocities
 ($V_{jet \hspace{0.1cm} \mathrm{ave}}$) and maximum jet velocities respectively as 
a function of $E_\mathrm{mg}/\dot{M_w}$. The sold line shows 
\[
V_{jet \hspace{0.1cm} \mathrm{ave}} \propto {(E_{\mathrm{mg}}/\dot{M_w})}^{0.16}.
\]

\[
V_{jet \hspace{0.1cm} \mathrm{max}} \propto {(E_{\mathrm{mg}}/\dot{M_w})}^{0.17}.
\]

It is interesting that (in case of week magnetic field) this relation shows an 
approximate agreement with the scaling law of the Michel's solution (\cite{Mic1969};
 \cite{BeM1976}); 
i.e., $v_{\infty} = {({\Phi}^2 {\Omega_F}^2 \big/ \dot{M_w})}^{1/3}$, 

where $v_{\infty}$ is the poloidal velocity at infinity and 
$\Phi = {r_0}^2 {B_0}$ and $\Omega_F$ is 
proportional to the Keplerian angular velocity of the disk. 

Figure \ref{13}a shows the time averaged mass accretion rates of the disk as 
a function of the initial $E_{\mathrm{mg}}$. 
The figure shows that its 
dependence on $E_{\mathrm{mg}}$ is about 
\[
\dot{M_a} \propto {E_{\mathrm{mg}}}^{0.6}.
\]

Figure \ref{13}b shows the the time averaged mass ejection rates of our jets 
as a function of the initial $E_{\mathrm{mg}}$. 
Its dependence on $E_{\mathrm{mg}}$ is approximately 
\[
\dot{M_w} \propto {E_{\mathrm{mg}}}^{0.16}.
\]

Figure \ref{13}c shows the the time averaged toroidal magnetic energy 
as a function of the initial $E_{\mathrm{mg}}$. 
Its dependence on $E_{\mathrm{mg}}$ is approximately 
\[
E_{\mathrm{mgt}} \propto {E_{\mathrm{mg}}}^{0.5}.
\]

Figure \ref{13}d shows the ratio of the time averaged mass outflow rate 
of the jet to the time averaged mass accretion rate of the disk 
as a function of the initial $E_{\mathrm{mg}}$. 
$\dot{M_w}/\dot{M_a}$ is of order of 0.1 or less 
\[
\dot{M_w}/\dot{M_a}\propto {E_{\mathrm{mg}}}^{-0.4}.
\]
 
%#######################################################
\subsection{Long time evolution of jet velocity}
%#######################################################
Figure \ref{14} shows the long time evolution of jet velocity which is
 calculated through
 the jet region $r=0-1$. Figure \ref{14}(a) shows the averaged jet velocity
 and 
Figure \ref{14}(b) shows the maximum jet velocity. Both the averaged jet
 velocity and maximum jet velocity never reach steady state and both of them
 have the
 same character of the variation. In case of strong initial magnetic field 
the jet velocity has the maximum value earlier than the weak initial magnetic
 filed case, but after the first 5 orbits the jet velocity becomes similar
 for all initial magnetic field. After the first 10 orbits the jet velocity 
for the initial weak magnetic field becomes higher than that in case of strong
 initial magnetic field case.

We notice that in all models the jet velocity decreases severely after the
 first
three orbits. We think that the high velocity of the jet in the first stage of 
evolution is result from both the effect of initial and boundary condition.
 Also, at the first stage of the simulation the position of the ejection point
 of the jet is near the gravitational center. With the simulation going on the 
density and pressure of the central region of disk increases which leads to
the decrease or stopping the accretion within that region. As a result for that
the position of the ejection point of the jet becomes far from the gravitational
 center so that the jet velocity decreases. Also, The effect of softening
 gravitational
  potential near the gravitational center leads to the decrease of the jet velocity.
 While the simulation goings on, the magnetic field lines are trapped inside the 
softening gravitational potential. Consequently, they lose their angular momentum
 and their rotational velocities become very small which leads to the decrease of
 the jet velocity.
      
%################################################
\subsection{Jet driving forces}
%###############################################
Fig. \ref{15} shows the time evolution of powers of the jet i.e. Poynting
 flux $F_{p,j}$, kinetic flux $F_{k,j}$ and enthalpy flux $F_{en,j}$ which are
 described as:  
\begin{equation}
F_{p,j} = -\int_{0}^{1}2\pi r\frac{c}{4\pi} (E \times B)_z dr 
\hspace{1cm} 
\mathrm{at\hspace{0.2cm}z} = 4, 
\end{equation}

\begin{equation}
F_{k,j}=\int_{0}^{1} 2\pi r \rho v^2 v_z  dr 
\hspace{2.45cm} 
\mathrm{at\hspace{0.2cm}z} = 4, 
\end{equation}

\begin{equation}
F_{en,j} =\int_{0}^{1} 2\pi r\rho v_z \Biggl(\frac{\gamma}{\gamma-1}\frac{p}{\rho}\Biggl)dr 
\hspace{1cm} 
\mathrm{at\hspace{0.2cm}z} = 4, 
\end{equation}
where $z$ is z-coordinates for jets to pass through. 

The Poynting flux $F_{p,j}$ plays the 
dominant effect in driving and accelerating the jet at the initial 
evolution of our simulation of a strong magnetic field case, (model 1)
 $E_{mg}=2\times 10^{-4}$ unit $t=50$, after that it is suddenly becoming
 very weak.

In case of initial weak magnetic field, model 5, $E_{mg}=5\times 10^{-6}$,
 during the initial stage of the evolution, the enthalpy flux is the dominant 
one until $t\sim 50$, and both kinetic flux and Poynting flux have nearly 
the same value. After $t\sim 50$, the dominant energy flux is the Poynting
 flux and the other two fluxes have nearly the same value. At the last stage 
of the evolution the Poynting flux decreases again and the enthalpy
 flux becomes the dominant one. In case of very weak initial magnetic field 
, model 9, $E_{mg}=2\times 10^{-7}$, the enthalpy flux is the dominant one 
until the last stage of the simulation except after $t\sim 225$ it decreases
 sharply.

Figure \ref{16} shows the relation between  the average enthalpy
 flux $F_{en,j \hspace{0.1cm} \mathrm{ave}}$, kinetic
 flux $F_{k,j \hspace{0.1cm} \mathrm{ave}}$
and Poynting flux $F_{p,j \hspace{0.1cm} \mathrm{ave}}$
 and  the initial magnetic field strength. The average is defined as;
 
\begin{equation}
F_{p,j \hspace{0.1cm} \mathrm{ave}} = -\int_{0}^{T_e}\int_{0}^{1}2\pi r\frac{c}{4\pi}  (E \times B)_z dr  dt 
\bigg/ 10 T_{crit}
\hspace{1cm} 
\mathrm{at\hspace{0.2cm}z} = 4, \label{32}
\end{equation}
%where $z_p$ is z-coordinates for jets to pass through. 

\begin{equation}
F_{k,j \hspace{0.1cm} \mathrm{ave}}=\int_{0}^{T_e}\int_{0}^{1} 2\pi r \rho v^2 v_z  dr dt 
\bigg/ 10 T_{crit} 
\hspace{2.45cm} 
\mathrm{at\hspace{0.2cm}z} = 4, \label{33}
\end{equation}

\begin{equation}
F_{en,j \hspace{0.1cm} \mathrm{ave}} =\int_{0}^{T_e}\int_{0}^{1} 2\pi r\rho v_z \Biggl(\frac{\gamma}{\gamma-1}\frac{p}{\rho}\Biggl)dr dt 
\bigg/ 10 T_{crit} 
\hspace{1cm} 
\mathrm{at\hspace{0.2cm}z} = 4, \label{34}
\end{equation}
where $T_e$ is the end of the simulation time and $ 10 T_{crit}$ is the sum
 of the time over which  the flux of the jet is averaged.
Figure \ref{8} shows that the averaged enthalpy flux have the same value
 whatever is the initial magnetic field strength, whereas both
 the averaged kinetic flux and averaged Poynting flux increase with increasing 
the initial magnetic field.

Kudoh \& Shibata (1997a) showed that the dominant energy of a jet depends on 
the strength of the magnetic field. When the poloidal component of the
 magnetic field is $B_p \propto r^{-2}$, the fast magnetosonic point appears
 far from the Alfv\'en point and the dominant energy of the jet is the Poynting
 flux. In our simulations, the initial magnetic field is uniform. In such
 models, the fast magnetosonic point locates far from the Alfv\'en point
 (\cite{Kuw2000}).

%########################################
\section{Radial jet Structure}
%#######################################
Figures \ref{17}, \ref{18} and \ref{19} show the radial dependence of mass
 and energy flux of
 the jet. The mass flux definition is described by equation \ref{18}, the toroidal
 magnetic energy is described by equation \ref{19}  and the Poynting flux,
  kinetic flux 
and enthalpy flux are described by equations \ref{32}-\ref{34} respectively.
 Figure \ref{17} show 
the radial
 profile of the density, poloidal velocity and toroidal field.  The radial
 dependence of density shows that the peak density (which defines the jet) in
 models 
1 and 4 (stronger initial magnetic field cases) is greater than that of model 6
(a weaker initial magnetic field case).
 The radial dependence of the poloidal velocity shows that the collimated flows
 show higher velocities closer to the disk axis. The radial dependence of the 
toroidal magnetic field shows that the maximum value is close to the disk axis 
 as expected for collimated jet. Figure \ref{18} shows the radial profile of the 
Poynting, enthalpy and kinetic flux. We plot the time
 averaged flux in annular rings around the symmetry axis (z-axis) to show the
 radial dependence of these energy flux. We show these radial dependence also
 by taking different initial magnetic field strength to show its effect.
 In case of model 4 the collimation of both mass flux and kinetic flux 
is clear. In figure \ref{18} we calculated the flux at fixed height (z=4). Next we 
will calculate the flux at different heights. At each height we calculate 
the position of the maximum of each energy flux. Figure \ref{19} shows the
 r-coordinate versus of z-coordinate of the maximum of mass flux, toroidal energy 
and kinetic flux for models 1, and 3. This figure shows 
how the maximum of each flux at different hight progresses in z-direction.
 From this figure we can notice that the collimation degree is different with
 the different kind of the flux and the initial magnetic field. In case of
 strong initial magnetic field the collimation is achieved earlier than weak
 case.
 Also the collimation is achieved after some height over the accretion disk. 

%#########################################
\section{Summary and discussion}
%#########################################
In this paper we have shown that long term 2.5-dimensional MHD numerical
 simulation
 of magnetized accretion flows leads to an intermittent jet like outflow which 
never reach steady state. Our simulation is an extension of the works of 
Matusmoto et al. (1996) and Kudoh et al. (1998). Both of them performed
 time-dependent 2.5-dimensional MHD numerical simulations of jets from
 accretion disks including the dynamical processes within the disk. They
 showed that the ejection mechanism of the jets is the same as that in steady 
theory; i.e. the centrifugal force along the poloidal field line
 accelerates the jets within an Alfv\'en radius and, above the Alfv\'en radius,
the jet is accelerated by magnetic pressure. The simulations for both 
of them last only for one inner disk rotation. What does happen for the
 characteristics of the jets if the simulations become long?. The answer of 
this question is given in this paper. 

%####################################################  
\subsection{The Physical Meaning of Time Evolution for
 $\dot{M}_w$, $\dot{M}_a$, and ${E}_{mgt}$}
%####################################################
Cosmic jets are ubiquitous, being quite often associated with new-born stars,
 X-ray binaries, active galactic nuclei and gamma-ray bursts. In all such cases
, jets and disks seem to be inter-related. Not only jets need disks in order to
 provide them with the ejected plasma and magnetic fields, but also disks
 need jets in order that the accreted plasma gets rid of its excess angular
 momentum to accrete. Observationally, there has already been accumulated 
enough evidence for such a correlation. For example, in star forming regions 
an apparent correlation is found between accretion diagnostics and outflow
 signatures (\cite{Har1995}). Hence, our current
 understanding is that jets are fed by the material of an accretion disk
 surrounding the central object. The mean controller between the mass
 accretion 
 rate $\dot{M}_a$ and the mass ejection rate $\dot{M}_w$ is the magnetic field 
 strength. Matusmoto et al. (1996) studied the dependence on the initial 
magnetic field strength. They showed that the ratio of the mass ejected as jet
 to the total mass of the flux tube was about 10\% of the accreting
 mass. Figure \ref{5} and figure \ref{11} show that time evolution of the 
mass outflow and toroidal magnetic field outflow is quasi-periodic and the
 periodicity of the jet can be related to the time needed for the initial
 magnetic field to be twisted to generate toroidal field. Sato et al. (2003) 
found that these periodicities are  around 2$\pi$. The toroidal magnetic field
 energy increases as magnetic field line is twisted and accumulated, 
because the magnetic field line is dragged with infalling and rotating gases.
Then the piled up energy is released by magnetically driven outflow that is
 triggered by magnetic reconnection so that the mass outflow is intermittent
 and has some periodicities.

Figure \ref{13}d show that the average $\dot{M}_w$ and  $\dot{M}_a$ (averaged over 
10 rotations) is closely related. The magnetic field strength is the controller
 of the relation between $\dot{M}_w$ and  $\dot{M}_a$. When the initial
 magnetic energy is weak the ratio between $\dot{M}_w$ and  $\dot{M}_a$ is
 high. The ratio decreasing with increasing the initial magnetic field.
  Pelletier \& Pudritz (1992) stated that if the disk magnetic field is reduced,
 there is a very large increase in the mass-loss rate in the wind. The point is
 that since a slower wind is driven in weaker disk fields, one must provide 
much more mass in order to carry off the same amount of disk angular momentum.
 At strong initial magnetic energy field case the ratio reach a constant
 value. These dependences are consistent with the results of the steady solution
 (\cite{KuS1997a}).

%##############################################################
\subsection{The Dependences on the Initial Magnetic Field Strength in
 the Weak field Case}
%############################################################
Kudoh \& Shibata (1997a) derived the dependence of the mass outflow rate ($\dot{M}_w$) 
and mass accretion rate ($\dot{M}_a$) on the initial magnetic field strength 
using a semi-analytical method, 
\begin{equation}
\dot{M}_w \propto {E_{\mathrm{mg}}}^{1/2} \hspace{0.2cm},
\end{equation}
\begin{equation}
\dot{M}_a \propto E_{\mathrm{mg}} \hspace{0.2cm}.
\end{equation}
Furthermore, Michel's scaling law (\cite{Mic1969}) is written as 
\begin{equation}
V_{\mathrm{jet}} \sim {\biggl( \frac{{\Phi}^2 {{\Psi}_F}^2}{\dot{M}_w}
 \biggr)}^{1/3} \propto \biggl( {\frac{E_{\mathrm{mg}}}{\dot{M}_w}}\biggr)^{1/3}\hspace{0.3cm}.
\end{equation}
On the other hand, we obtain 
\begin{equation}
\frac{\dot{M}_{w,\mathrm{ave}}}{\dot{M}_{a,\mathrm{ave}}} \propto {E_\mathrm{mg}}^{-0.4}
 \hspace{0.2cm},
\end{equation}

\begin{equation}
V_{jet,\mathrm{ave}} \propto E_{\mathrm{mg}}^{0.15} \hspace{0.2cm}. 
\end{equation}

Our results shown in Figure \ref{12} and Figure \ref{13} are consistent with not
 only the steady solutions 
but also the results of both Kudoh et al. (1998), Kato et al. (2002) and 
Kigure \& Shibata (2005)
 However there is a remarkable difference from the results of Kudoh et al.
 (1998), Kato et al. (2002) and Kigure \& Shibata (2005).
 They show only maximum values for $\dot{M}_w$,
 $\dot{M}_a$ and $V_z$, while we examined time-averaged values. We find that
 $V_{z \hspace{0.1cm} \mathrm{ave}}$ and $\dot{M}_w$ are $0.1$ times smaller 
than $V_{z \hspace{0.1cm} \mathrm{max}}$
 and $\dot{M}_{w \hspace{0.1cm} \mathrm{max}}$ in the case of Kudoh et al.
 (1998). 
%##################################################
\section{Conclusions}
%#################################################
We have performed time-dependent, 2.5-dimensional, 
axially symmetric, MHD numerical simulations of jets for many orbital periods. 
Our simulations solve the dynamical 
process of the accretion disk for longer time case than that of the previous
 simulations. Because we wanted to know whether the jet continued to be
 ejected and become steady (or quasi-steady) state, we investigated time
 variations of $\dot{M}_w$, $\dot{M}_a$, and $E_{mgt}$. 
We also wanted to know the dependence of $\dot{M}_{w \hspace{0.1cm}
 \mathrm{ave}}$, $\dot{M}_{a \hspace{0.1cm} \mathrm{ave}}$, and
 $V_{z \hspace{0.1cm} \mathrm{ave}}$ on the initial magnetic field strength
 and compared our results with the steady theory and cases of Kudoh et al.
 (1998) and Kato et al. (2002). We summarize our results as follows: 

1. In all models, the ejection of jets is intermittent. 

2. The ejection of jets has a period which related to the toroidal
 field formation. 

There are also relationships between $\dot{M}_w$ and $\dot{M}_a$, 
and between $\dot{M}_w$ and $E_{mgt}$; specifically, their  time variations 
are similar. 

3. The dependence of $\dot{M}_w$, $\dot{M}_a$, and $E_{mgt}$ on the strength
 of 
the initial magnetic field is consistent with those in the steady theory 
and cases of  Kudoh et al. (1998) and Kato et al. (2002).
In all cases, however, $V_{z \hspace{0.1cm} \mathrm{ave}}$ and $\dot{M}_w$
 are $0.1$ times smaller 
than $V_{z \hspace{0.1cm} \mathrm{max}}$ and $\dot{M}_{w \hspace{0.1cm} 
\mathrm{max}}$ 
in the case of Kudoh et al. (1998) and Kato et al. (2002) and Kigure \& Shibata (2005)

%#######################################
\noindent  Acknowledgment
%######################################

This work is supported by the Grant-in-Aid for the
21st Century COE ``Center for Diversity and Universality in
Physics'' from the Ministry of Education, Culture, Sports, Science
and Technology (MEXT) of Japan. Numerical computations
were carried out on VPP5000 at the Astronomical Data
Analysis Center of the National Astronomical Observatory,
Japan (project ID: iaa31c), an inter-university research
institute of astronomy operated by the Ministry of Education,
Culture, Sports, Science, and Technology.

\include{table}
\include{figures}

\end{document}

%% file: table.tex
\begin{table}[htbp]
\begin{center}
\begin{tabular}{ccc}
\multicolumn{3}{c}{\textbf{TABLE 1}} \\
\multicolumn{3}{c}{} \\
\multicolumn{3}{c}{\textbf{UNITS FOR NORMALIZATION}} \\
\multicolumn{3}{c}{} \\
\hline \hline
\textbf{}	&	\textbf{Normalization}		&	\textbf{}	\\
\textbf{Physical Quantities}	&	\textbf{Unit}		&	\textbf{}	\\
\hline
$t$(time) $\cdots\cdots\cdots\cdots\cdots\cdots\cdots\cdots\cdot$		&	$r_0/V_{\mathrm{K0}}$	&	\\

$r,z$(length) $\cdots\cdots\cdots\cdots\cdots\cdots\cdots$		&	$r_0$		&	\\

$\rho$(density) $\cdots\cdots\cdots\cdots\cdots\cdots\cdots\cdot$		&	$\rho_0$	&	\\

$p$(pressure)	$\cdots\cdots\cdots\cdots\cdots\cdots\cdots$	&	$\rho_0V^2_{\mathrm{K0}}$	&	\\

$v$(velocity)	$\cdots\cdots\cdots\cdots\cdots\cdots\cdots\cdot$		&	$V_{\mathrm{K0}}$	&		\\

$B$(magnetic field) $\cdots\cdots\cdots\cdots\cdots$	&	$(\rho_0V^2_{\mathrm{K0}})^{1/2}$	&		\\

$e$(specific internal energy)	$\cdots\cdots\cdot$	&	$V^2_{\mathrm{K0}}$		&		\\
\hline \hline 
%\multicolumn{3}{c}{The unit length $r_0=(L^2_0/GM)^(1/(1-2a)$ is the radius of}\\
%\multicolumn{3}{c}{the density maximum in the initial disk. The unit velocity}\\
%\multicolumn{3}{c}{$V_{K} \equiv (GM/r_0)^{(1/2)}$ is the Keplerian velocity at $(r,z)=(r_0,0)$.}\\
%\multicolumn{3}{c}{The unit density $\rho_0$ is the initial density at $(r,z)=(r_0,0)$. It}\\
%\multicolumn{3}{c}{is assumed that $a=0.45$, in our study (see eq. 12)}
\end{tabular}
\caption{The unit length $r_0=(L^2_0/GM)^(1/(1-2a)$ is the radius 
of the density maximum in the initial disk. The unit velocity $V_{K} \equiv (GM/r_0)^(1/2)$ 
is the Keplerian velocity at $(r,z)=(r_0,0)$. The unit density $\rho_0$ is the initial density 
at $(r,z)=(r_0,0)$. It is assumed that $a=0.45$, in our study (see eq.\ref{12} for the definition of a)}
\end{center}
\end{table}

\begin{table}[htbp]
\begin{center}
\begin{tabular}{cccccccc}
\multicolumn{8}{c}{\textbf{TABLE 2}} \\
\multicolumn{8}{c}{} \\
\multicolumn{8}{c}{\textbf{MODEL PARAMETERS}} \\
\multicolumn{8}{c}{} \\
\hline \hline
\textbf{model}	&&	\textbf{$E_{\mathrm{mg}}$}	&&	\textbf{$E_{th}$}	&&	\textbf{time}	&	\textbf{}	\\
\hline
1				&&	$2\times10^{-4}$			&&	0.018				&&	$80.0\pi$					\\

2				&&	$8\times10^{-5}$			&&	0.018				&&	$22.1\pi$					\\

3				&&	$5\times10^{-5}$			&&	0.018				&&	$31.7\pi$					\\

4				&&	$2\times10^{-5}$			&&	0.018				&&	$31.2\pi$					\\
5				&&	$5\times10^{-6}$			&&	0.018				&&	$80.0\pi$					\\

6				&&	$2\times10^{-6}$			&&	0.018				&&	$37.7\pi$					\\

7				&&	$8\times10^{-7}$			&&	0.018				&&	$41.3\pi$					\\

8				&&	$5\times10^{-7}$			&&	0.018				&&	$34.8\pi$					\\
9				&&	$2\times10^{-7}$			&&	0.018				&&	$80.0\pi$					\\
\hline \hline
\end{tabular}
%\caption[]{}
\label{}
\end{center}
\end{table}

%% file: figures.tex
\begin{figure}
  \begin{center}
    \FigureFile(168mm,168mm){fig1.eps}
  \end{center}
  \caption{Temporal evolution of an ideal accretion disk threaded by a 
poloidal magnetic field. Color levels represent density level (upper panel)
 and temperature level (bottom panels) while solid lines stand for poloidal 
magnetic filed
lines. The time unit labeling each snapshot is for $t\sim 2\pi \sim 6.28$ 
corresponds to one Keplerian orbit at $(r,z)= (1,0)$. After a few rotations,
 outflows are escaping from the disk and the outflow and accretion remain
 until the last moment of our simulation time. The  square refers to the region
 analyzed by Kudoh et al 1998 }\label{1}
\end{figure}

\begin{figure}
  \begin{center}
    \FigureFile(168mm,168mm){fig2.eps}
  \end{center}
  \caption{Temporal evolution of an ideal accretion disk threaded by a 
poloidal magnetic field. Color levels represent density level (upper panel)
 and temperature level (lower panel) while solid lines stand for poloidal
 magnetic filed
lines. The time unit labeling each snapshot is for $t\sim 2\pi \sim 6.28$
 corresponds to one Keplerian orbit at $(r,z)= (1,0)$. After a few rotations,
 outflows are escaping from the disk and the outflow and accretion remain 
until the last moment of our simulation time }\label{2}
\end{figure}

\begin{figure}
 \begin{center}
    \FigureFile(168mm,168mm){fig3.eps}
   \end{center}
  \caption{Temporal evolution of an ideal accretion disk threaded by a 
poloidal magnetic field. Color levels represent density level (upper panel) 
and temperature level (bottom panels) while solid lines stand for poloidal 
magnetic filed
lines. The time unit labeling each snapshot is for $t\sim 2\pi \sim 6.28$ 
corresponds to one Keplerian orbit at $(r,z)= (1,0)$. After a few rotations,
 outflows are escaping from the disk and the outflow and accretion remain 
until the last moment of our simulation time}\label{3}
\end {figure}

\begin {figure}
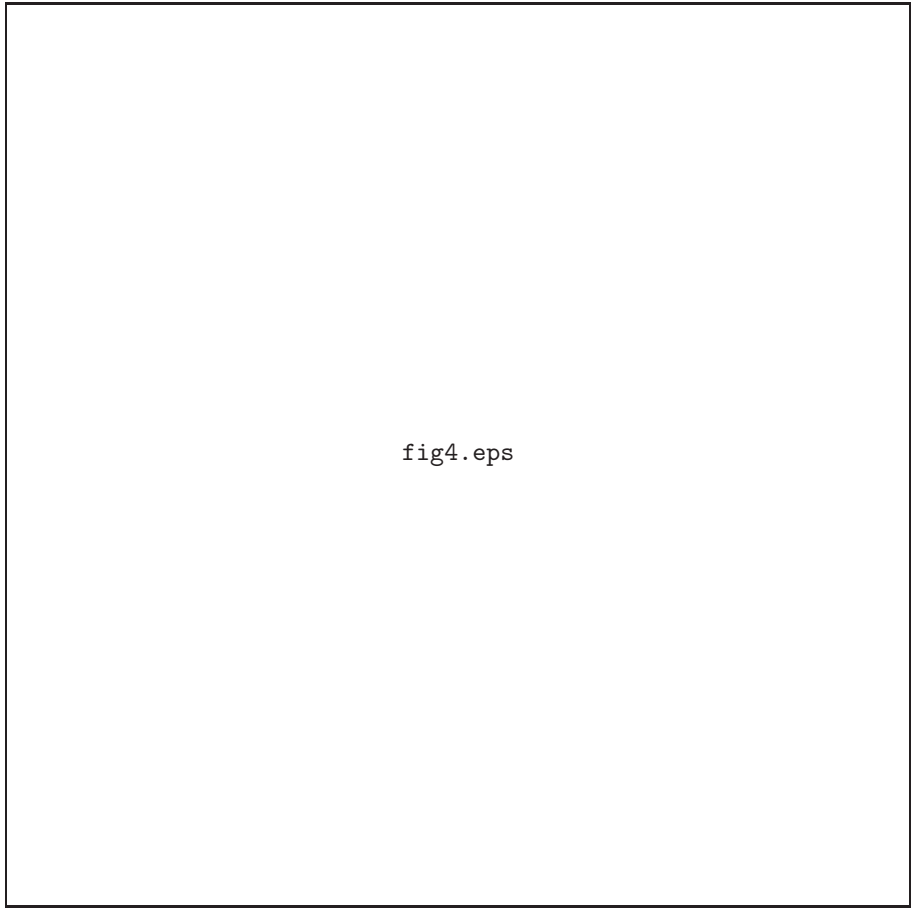

\begin{center}
    \FigureFile(120mm,120mm){fig4.eps}
  \end{center}
  \caption{Ratio of $B_\varphi$ to $B_p$ along  $r= 0.225 $. at different three
time shot, for different magnetic field energy. The twisting magnetic field 
become more prominent in case of weak magnetic field.}\label{4}
\end {figure}

%##########Mass Flux#######################

\begin{figure}
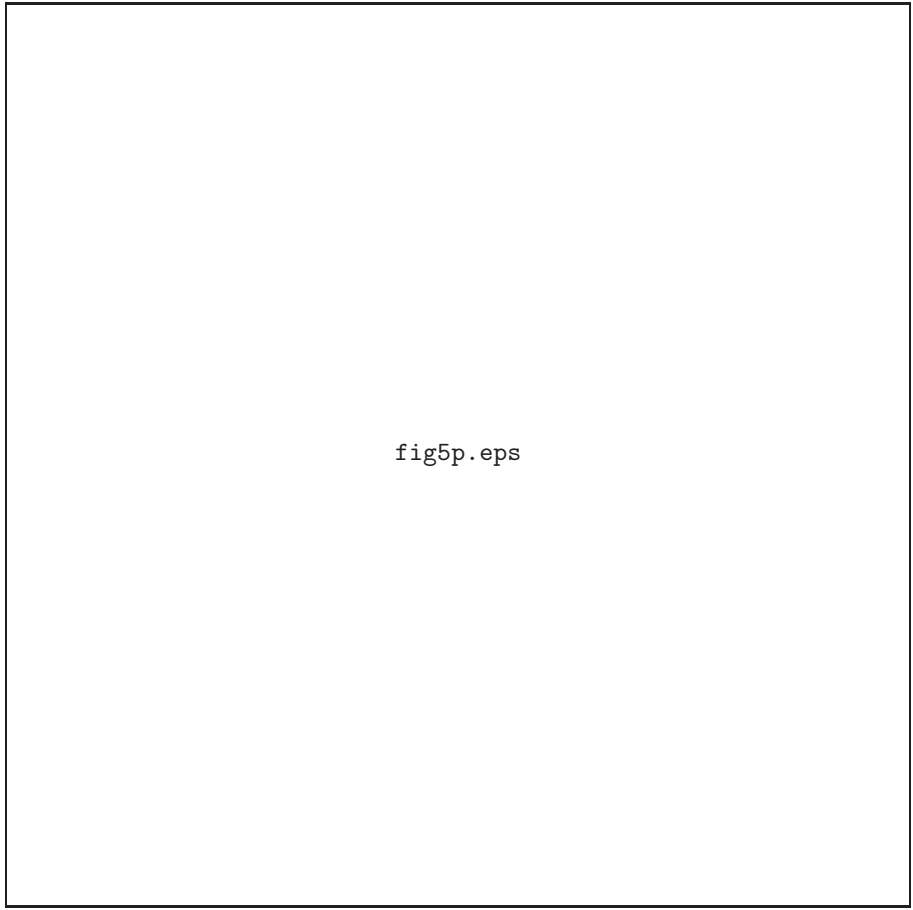

  \begin{center}
    \FigureFile(120mm,120mm){fig5p.eps}
  \end{center}
  \caption{Time variation of the mass outflow rate for three models, the out
 flow is still intermittent after long time simulation for different initial
magnetic field energy }\label{5}
\end{figure}

\begin{figure}
  \begin{center}
    \FigureFile(168mm,168mm){fig6.eps}
  \end{center}
  \caption{Time variation of the mass accretion  rate for three models, the 
accretion is still intermittent after long time simulation for different
 initial magnetic field energy. One Keplerian orbit at $(r,z)= (1,0) \sim 2\pi \sim 6.28$ time  unit.}\label{6}
\end{figure}

%############### Mass accretion as function of time and radius ######

\begin{figure}
  \begin{center}
    \FigureFile(168mm,168mm){fig7.eps}
  \end{center}
  \caption{ Accretion rate, averaged between $r=0.52$ and $r=10$, against time
 for different initial magnetic field energy. One Keplerian orbit at $(r,z)= (1,0) \sim 2\pi \sim 6.28$ time unit}\label{7}
\end{figure}

\begin{figure}
  \begin{center}
    \FigureFile(168mm,168mm){fig8.eps}
  \end{center}
  \caption{ Accretion rate as a function of radius, for different initial
 magnetic field, averaged over 10 orbites for models 1, 3, 4 and 6}\label{8}
\end{figure}
 
\begin{figure}
  \begin{center}
    \FigureFile(168mm,168mm){fig9.eps}
  \end{center}
  \caption{Time variation of the toroidal energy  for three models, the outflow
 is still intermittent, similar to mass out flow, after long time
 simulation for different initial magnetic field energy. One Keplerian orbit at $(r,z)= (1,0) \sim 2\pi \sim 6.28$ time unit}\label{9}
\end{figure}

\begin{figure}
  \begin{center}
    \FigureFile(168mm,168mm){fig10p.eps}
  \end{center}
  \caption{The mass flux for four models as a function of time. The numbers
 correspond to the maximum flux at which we measure the corresponding time 
(peak time ) and plot this time with the initial magnetic field in fig. \ref{11}}\label{10}
\end{figure}

\begin{figure}
  \begin{center}
    \FigureFile(168mm,168mm){fig11pp.eps}
  \end{center}
  \caption{In this figure we show the relation between the average time
 interval between the mass ejection peaks and the initial magnetic field
 strength (a)
 and the average magnetic  field (b). Solid line corresponds to the best fit to
 the numerical result and the dotted line corresponds to the analytical
 relation ( equation \ref{26}).}\label{11}
\end{figure}

%#########################velocity averaged relation######################

\begin{figure}
  \begin{center}
    \FigureFile(160mm,160mm){fig12.eps}
  \end{center}
  \caption{(a) Time averaged 
velocities ($V_{jet \hspace{0.1cm} \mathrm{ave}}$) as a function of
 $E_\mathrm{mg}$.(b) $V_{z \hspace{0.1cm} \mathrm{max}}$ as a function
 of $E_\mathrm{mg}$. (c) Time averaged velocities($V_{jet \hspace{0.1cm}
 \mathrm{ave}}$) as a function of $E_\mathrm{mg}/\dot{M_w}$.
 (d) $V_{jet \hspace{0.1cm} \mathrm{max}}$  as a function of 
$E_\mathrm{mg}/\dot{M_w}$.}\label{12}
\end{figure}

\begin{figure}
  \begin{center}
    \FigureFile(160mm,160mm){fig13.eps}
  \end{center}
  \caption{ (a) Time averaged mass accretion ($\dot{M_a}$) rate of the disk as
 a function of $E_{\mathrm{mg}}$. (b)  Time averaged mass ejection 
($\dot{M_w}$) rate of the jet as a function of $E_{\mathrm{mg}}$. (c) Average
 toroidal ejection ($E_{mgt}$) as a function of $E_{\mathrm{mg}}$. (d)
 The ratio of the time averaged mass outflow rate of the jet to the time
 averaged mass accretion rate of the disk ($\dot{M_w}$/$\dot{M_a}$) as a
 function of $E_{\mathrm{mg}}$.}\label{13}
\end{figure}

\begin {figure}
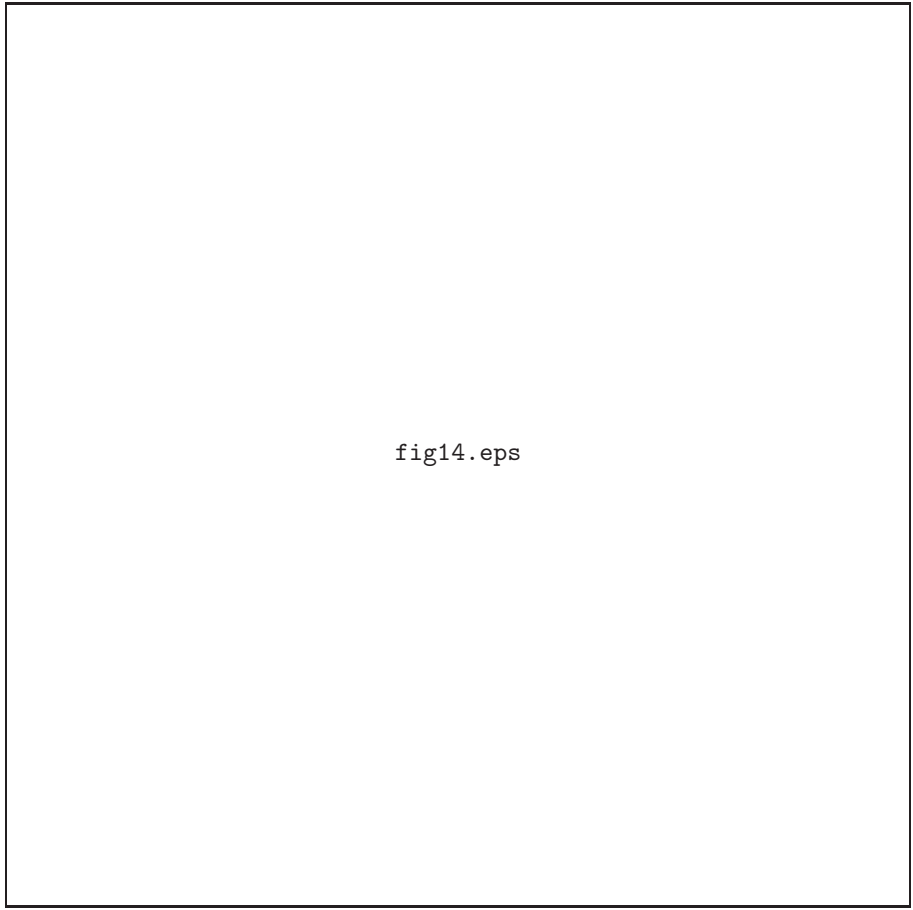

\begin{center}
    \FigureFile(120mm,120mm){fig14.eps}
  \end{center}
\begin{center}
  \caption{The jet velocity for three models as a function in time. (a)
 The averaged jet velocity (averaged between $r=0$ and $r=1$ ) at height
 $z=4$ as a function in time. (b) The maximum jet velocity at $z=4$ as a
 function in time (averaged between $r=0$ and $r=1$. One Keplerian orbit at $(r,z)= (1,0) \sim 2\pi \sim 6.28$ time unit}\label{14}
\end{center}
\end {figure}

\begin {figure}
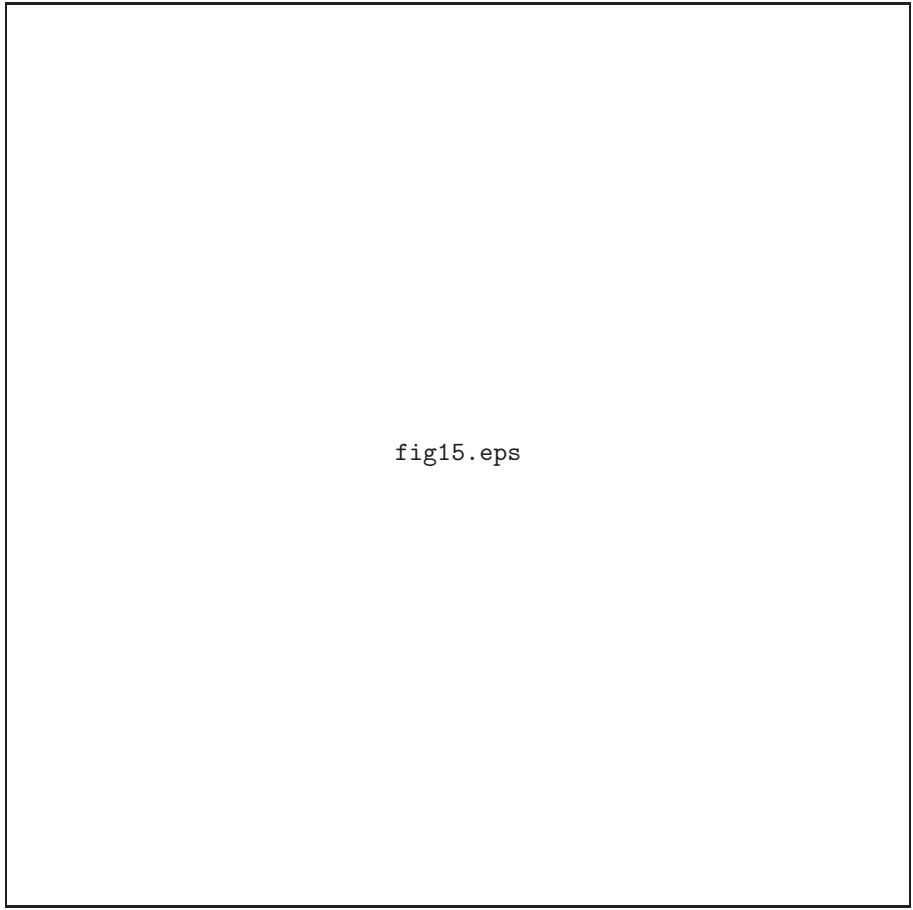

\begin{center}
    \FigureFile(120mm,120mm){fig15.eps}
  \end{center}
  \caption{Time evolution of powers of the jet; Poynting flux,
 $F_{p,j}$ (solid line), thermal flux $F_{en,j}$ (dashed line), and kinetic
 flux $F_{k,j}$ (dotted line) for different initial magnetic field}\label{15}
\end {figure}

\begin {figure}
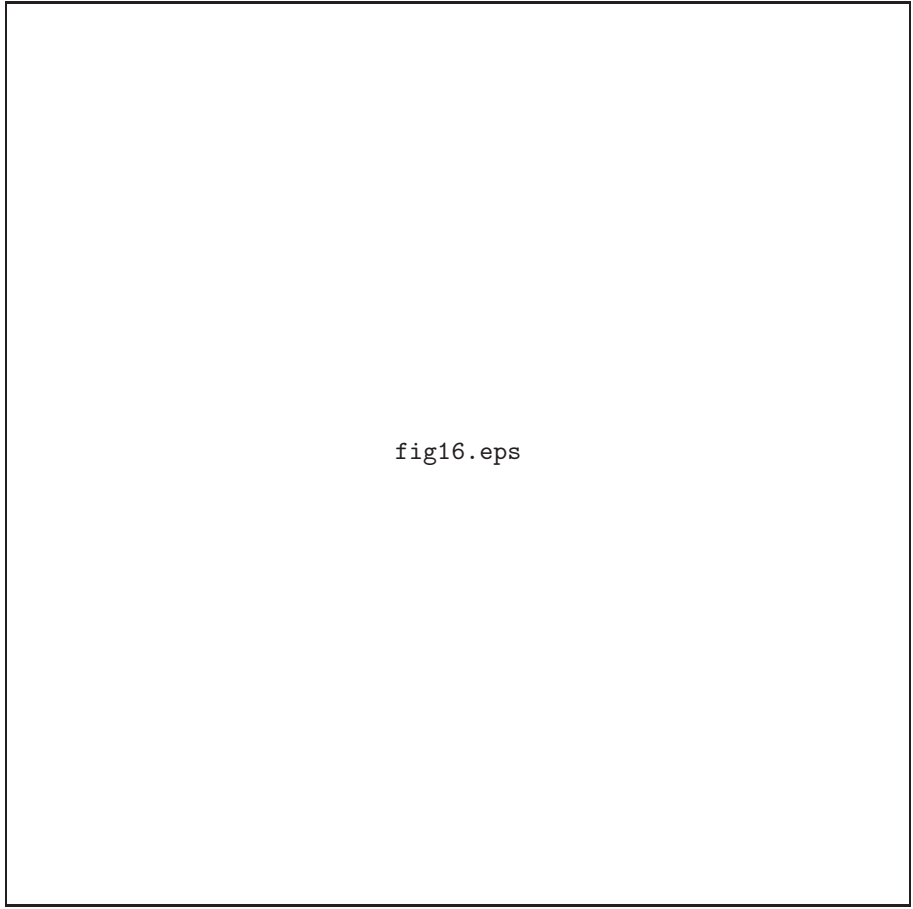

\begin{center}
    \FigureFile(120mm,120mm){fig16.eps}
  \end{center}
  \caption{The time average value of enthalpy, kinetic and Poynting flux as 
a relation with the initial magnetic field strength.}\label{16}
\end {figure}

\begin{figure}
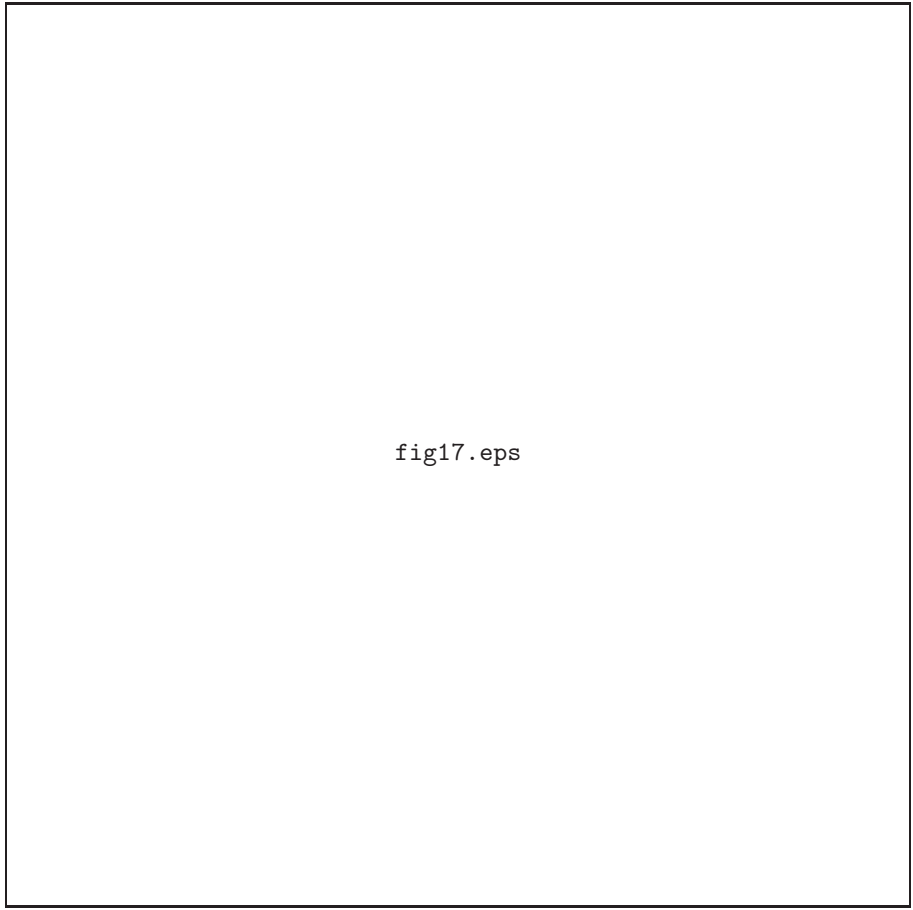

  \begin{center}
    \FigureFile(120mm,120mm){fig17.eps}
  \end{center}
  \caption{Radial profiles of average physical quantities. The cuts were taken
 at $z=4$. The radial dependence of density for each initial magnetic field
 shows that the peak density (which defines the jet) in the first two initial
 magnetic field is greater than the last weak initial magnetic field case.
 The radial dependence of the poloidal velocity shows that the collimated flows
 show higher velocities closer to the disk axis. The radial dependence of the 
toroidal magnetic field shows that the maximum value is close to the disk axis 
 as expected for collimated jet.}\label{17}
\end{figure}

\begin{figure}
  \begin{center}
    \FigureFile(168mm,168mm){fig18.eps}
  \end{center}
  \caption{This figure shows the flux outflow of (a) mass flux ,(b) kinetic
flux,(c) toroidal magnetic energy, (d) enthalpy flux, and (e) Poynting flux 
 at z=4 as a function of r-coordinate.
 The collimation becomes more clear in case of model 4 in both mass flux
 and kinetic flux }\label{18}
\end{figure}

\begin{figure}
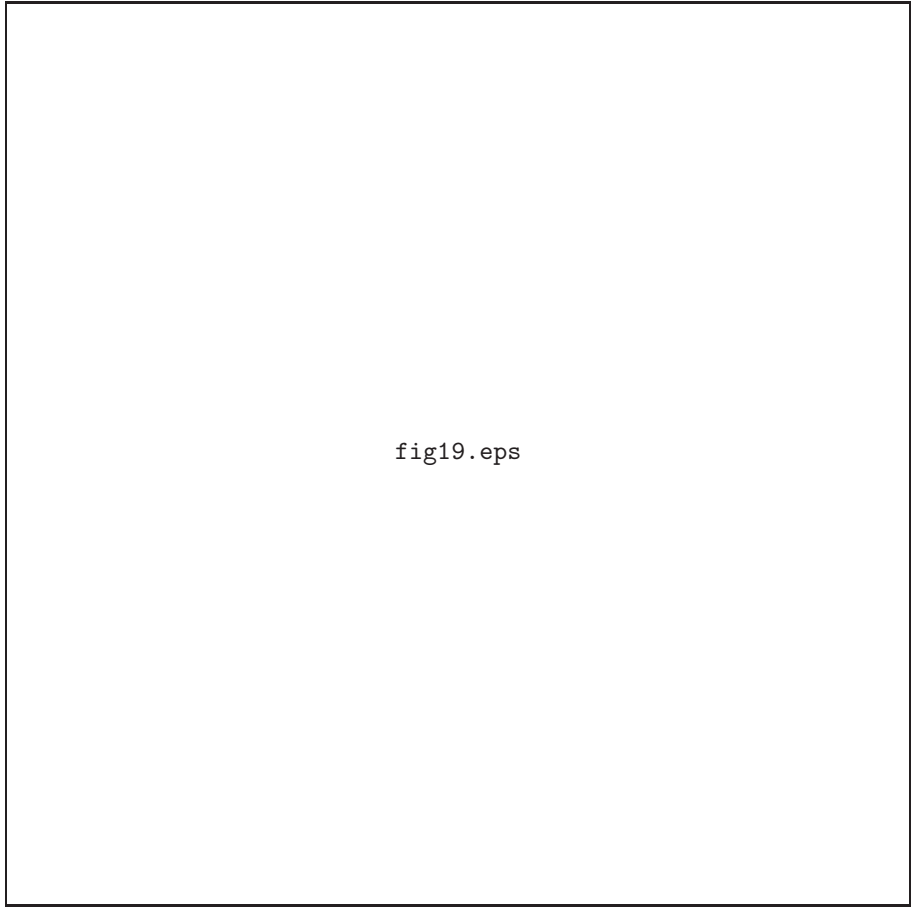

  \begin{center}
    \FigureFile(120mm,120mm){fig19.eps}
  \end{center}
  \caption{The lines represent the progress of the maximum of
 different fluxes in z-direction. The collimation begins at low-z in case
 of strong initial magnetic field.}\label{19}
\end{figure}